\documentclass[aps,twocolumn,superscriptaddress,prl]{revtex4}
\usepackage{amssymb}
\usepackage{amsmath}
\usepackage{graphicx}

\begin{document}
\title{Gate-Controlled Spin-Orbit Quantum Interference Effects in
Lateral Transport}
\author{J.\ B.\ Miller}
\affiliation{Department of Physics, Harvard University, Cambridge,
Massachusetts 02138}
\affiliation{Division of Engineering and Applied Science, Harvard
University, Cambridge, Massachusetts 02138}
\author{D.\ M.\ Zumb\"uhl}
\affiliation{Department of Physics, Harvard University, Cambridge,
Massachusetts 02138}
\author{C.\ M.\ Marcus}
\affiliation{Department of Physics, Harvard University, Cambridge,
Massachusetts 02138}
\author{Y.\ B.\ Lyanda-Geller}
\affiliation{Naval Research Laboratory, Washington, D.C. 20375}
\author{D.\ Goldhaber-Gordon}
\affiliation{Department of Physics, Harvard University, Cambridge,
Massachusetts 02138}
\affiliation{Department of Physics, Stanford University, Stanford,
California 94305}
\author{K.\ Campman}
\author{A.\ C. Gossard}
\affiliation{Materials Department, University of California at
Santa Barbara, Santa Barbara, California, 93106}

\begin{abstract}
\emph{In situ} control of spin-orbit coupling in coherent
transport using a clean GaAs/AlGaAs two-dimensional electron gas
is realized, leading to a gate-tunable crossover from weak
localization to antilocalization. The necessary theory of 2D
magnetotransport in the presence of spin-orbit coupling beyond the
diffusive approximation is developed and used to analyze
experimental data. With this theory the Rashba contribution and
linear and cubic Dresselhaus contributions to spin-orbit coupling
are separately estimated, allowing the angular dependence of
spin-orbit precession to be extracted at various gate voltages.
\end{abstract}
\maketitle
An important component along the path toward realizing quantum
``spintronic" devices \cite{Wolf,AwschalomBook} is a structure
that allows manipulation of electron spin without destroying phase
coherence. Spin-orbit (SO) coupling has been the focus of recent
studies because of its potentially useful role in coherent spin
rotators \cite{Datta90}, spin interference devices \cite{Aronov},
and spin-filters \cite{Koga,Kiselev}. The mechanisms by which SO
coupling affects transport
\cite{Hikami80,AALKh81,Bergmann,Dresselhaus92} have recently been
considered in the context of Aharonov-Bohm (AB) phase and Berry
phase
\cite{Aronov,Stone,Mirlin,Iordanskii94,Lyanda98,Aleiner,Zumbuhl},
underscoring the richness of the underlying physics.

The conductivity of low-dimensional systems shows signatures of
quantum interference that depend on magnetic field and SO coupling
\cite{Hikami80,AALKh81,Anderson,Altshuler1,Gorkov}. In particular,
constructive (destructive) backscattering associated with pairs of
time-reversed closed-loop electron trajectories in the absence
(presence) of significant SO interaction leads to negative
(positive) magnetoresistance effects known as weak localization
(antilocalization) \cite{Bergmann}.
In this Letter, we demonstrate \emph{in situ} control of SO
coupling in a moderately high mobility GaAs/AlGaAs two-dimensional
electron gas (2DEG), inducing a crossover from weak localization
(WL) to antilocalization (AL) as a function of an applied top-gate
voltage (see Fig.~1). Theory beyond the diffusive approximation
must be used to extract gate-voltage-dependent SO parameters from
magnetotransport when the SO precession frequency becomes
comparable to the inverse transport scattering time ($\tau^{-1}$)
as occurs here, and when the magnetic length becomes comparable to
the mean free path. Theory that accounts for AB-like spin phases
and spin-relaxation \cite{Lyanda1} is developed here and used to
estimate {\em separately} the various SO terms (Rashba, linear and
cubic Dresselhaus) over a range of gate voltages, ranging from WL
to AL.
Conventional WL theories assume SO times much longer than $\tau$
\cite{Hikami80,AALKh81,Iordanskii94} and so cannot be applied to
clean materials.  Previous theories that go beyond the diffusive
approximation do not treat SO \cite{Gasparyan85, REF:Shapiro}, or
treat it only as spin-relaxation \cite{Kawabata,Zduniak97} without
accounting for Berry phase effects which play a crucial role, as
we show here.
\begin{figure}[!tbp]
\center \label{fig1} \includegraphics[width=2.8in]{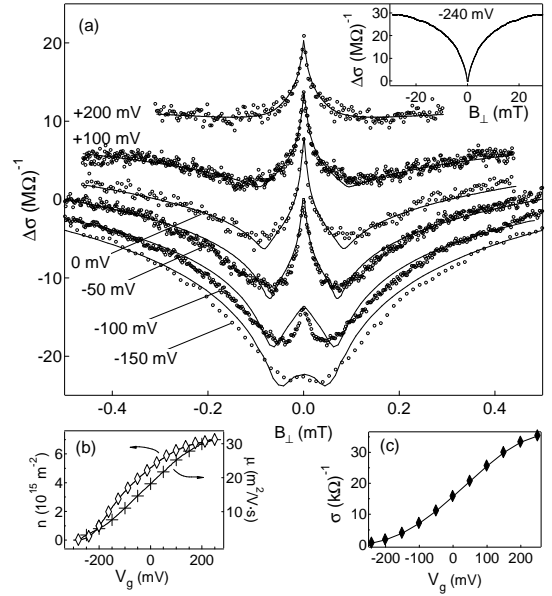}
\caption{\footnotesize {(a) Experimental magnetoconductance,
$\Delta\protect\sigma=\protect \sigma(B)-\protect\sigma(0)$,
(circles) offset for clarity, along with three-parameter fits to
Eq.~(\ref{Cond}) (solid curves) for several gate voltages. Inset:
Experimental magnetoconductance data for the most negative gate
voltage, showing pure WL. (b) Density and mobility as a function
of $V_g$, extracted from longitudinal and Hall voltage
measurements. (c) Experimental conductivity, showing strong
dependence on $V_g$. Note that  $\Delta\sigma \sim 10^{-3}
\sigma$.}}
\end{figure}
Previous experiments in which SO rates are measured using WL/AL in
a gated GaAs heterostructure have not reported \emph{in situ} gate
control \cite{Dresselhaus92, Taborski, Ramvall97}. Koga \emph{et
al.}\ \cite{Koga2} demonstrated gate controlled SO coupling in
InGaAs heterostructures using WL/AL. Modification of Rashba SO
coupling using gated quantum wells has been observed using beating
patterns in Shubnikov-de Haas oscillations in InGaAs
\cite{Nitta97,Schapers98}, InAs/AlSb \cite{Heida98} and HgTe
\cite{Schultz96}.  Gate controlled SO coupling in GaAs 2D hole
systems \cite{Papadakis01,Winkler02,Lu98} has also been
investigated using beating of Shubnikov-de Haas oscillations. The
angular variation of SO coupling in GaAs quantum wells has been
measured using Raman scattering \cite{Jusserand95}.

The Hamiltonian for conduction band electrons in a [001] 2DEG is
$\mathcal{H}=\frac{\hbar ^{2}k^{2}}{2m^*}+(\mathbf{{\sigma }\cdot
{\Omega })}$, where $m^*$ is the effective mass, $k=|\mathbf{k}|$
($\mathbf{k}=(k_x,k_y)$) is the in-plane wave vector,
$\mathbf{\sigma}=(\sigma _{x},\sigma _{y},\sigma_{z})$ is the
Pauli spin operator and $\mathbf{\Omega }=(\Omega _{x},\Omega
_{y})$ is the total SO frequency. $\mathbf{\Omega }$ can be
written as the vector sum of linear ($\mathbf{\Omega}_{D1}$) and
cubic ($\mathbf{ \Omega} _{D3}$) Dresselhaus terms and the Rashba
term ($\mathbf{\Omega} _{R}$),
\begin{subequations}
\label{Omegas}
\begin{eqnarray}
\mathbf{\Omega} _{D1} &=&\alpha _{1}\left( -\mathbf{\hat{x}}k_x+
\mathbf{\hat{y}}k_y\right)/\hbar, \\
\mathbf{\Omega}_{R} &=&\alpha
_{2}(\mathbf{\hat{x}}k_y-\mathbf{\hat{y}}
k_x)/\hbar, \\
\mathbf{\Omega}_{D3} &=&\gamma (\mathbf{\hat{x}}k_xk_
y^2-\mathbf{\hat{y}}k_x^2k_y)/\hbar. \label{Omega3Eqn}
\end{eqnarray}
\end{subequations}
\noindent where $\gamma$ arises from the lack of inversion
symmetry of the GaAs crystal, while $\alpha_1=\gamma\langle
k_{z}^{2}\rangle$ also depends on the thickness of the wave
function in the quantization direction. $\alpha_2$ depends on the
potential profile of the heterointerface. We assume the effect of
gate voltage, $V_{g}$, on $\Omega$ $(\equiv |\mathbf{\Omega}|)$ is
through the carrier density, $n = k^2/2\pi$. Previous studies of
SO coupling in single-interface heterostructures \cite{Knap96}
support this assumption. The magnitude of $\alpha_2$ in a
single-interface heterostructure originates mainly from the
band-offset at the heterointerface, which is roughly independent
of $V_g$ \cite{Heida98,Pfeffer99}.
The symmetry of the linear (in $k$) SO terms, $\mathbf{\Omega}
_{D1}$ and $\mathbf{\Omega} _{R}$, allows these terms to be
represented as a spin-dependent vector potential $\mathbf{A}$ that
affects the orbital motion and phase of electrons,
$\mathbf{\sigma}\cdot
(\mathbf{\Omega}_{D1}+\mathbf{\Omega}_{R})\propto \mathbf{k}\cdot
\mathbf{A}$
\cite{Aronov,Stone,Mirlin,Iordanskii94,Lyanda98,Aleiner}. That is,
the linear terms affect electronic interference as a
spin-dependent AB-like effect. In contrast, the cubic term,
Eq.~(1c), upon removing terms with the symmetry of Eq.~(1a), only
causes spin relaxation in the diffusive regime (although it also
can produce AB-like effects in the quasi-ballistic regime
\cite{Aronov}).
To develop the theory of 2D magnetotransport with SO coupling
beyond the diffusive approximation \cite{Lyanda1}, we follow
Refs.~\cite{Gasparyan85,REF:Shapiro,Kawabata}, which treat the
quasi-ballistic case $\ell_B < \ell$ ($\ell_B = \sqrt{\hbar /2eB}$
is the magnetic length and $\ell$ is the transport mean free path)
without SO coupling. The approach is to introduce an operator
$P=G_{\epsilon +\omega }^{R}(\mathbf{
r}_{1},\mathbf{r}_{2},\mathbf{\sigma} _{1})G_{\epsilon
}^{A}(\mathbf{r}_{1}, \mathbf{r}_{2},\mathbf{\sigma} _{2})\hbar
/2\pi \nu \tau$ for the probability of an electron to propagate
both forward and backward along a path segment from
$\mathbf{r}_{1}$ to $\mathbf{r}_{2}$, where $G^{R}$ ($G^{A}$) are
single-electron retarded (advanced) Green functions,
$\mathbf{\sigma}_{1(2)}$ are the Pauli spin operators for particle
moving forward (backward), $\nu$ is the density of states per
spin, and $\tau$ is the scattering time. The interference
contribution from traversing a closed trajectory with $n$
scattering events is given by the trace of $(P)^n$. In the
presence of SO coupling, Eq.~(\ref{Omegas}), the formulas in
\cite{REF:Shapiro} remain valid once a summation over spins is
included in the trace.
Introducing the total spin of interfering particle waves,
$\mathbf{S}=\mathbf{\sigma}_{1}+\mathbf{\sigma}_{2}$, we write
$Tr[(P)^{n}]=\frac{1}{2} Tr[(P_{1})^{n}-(P_{0})^{n}]$, where
operators $P_{0}$ and $P_{1}$ describe  singlet ($S=0$) and
triplet ($S=1$) contributions.  To calculate $Tr[(P_{0(1)})^{n}]$,
we diagonalize $P_{0(1)}$. We find that when
$\mathbf{\Omega}_{D1}$ and $\mathbf{\Omega}_{R}$ are taken into
account, $P_{0(1)}$ has the same eigenfunctions as the Hamiltonian
$\mathcal{H}$ for particles with charge 2$e$, spin $\mathbf{S}$
and spin frequency $2\mathbf{\Omega }$: $\mathcal{H} =
\frac{\hbar^{2}}{2m^*}(\mathbf{k}-2e\mathbf{A}_{em}+
2e\mathbf{A}_{S})^{2}$, where $\mathbf{A}_{em}$ is the vector
potential associated with the applied perpendicular magnetic
field, $B$, and $\mathbf{A}_{S}=\frac{m^*}{2e\hbar ^{3}}(-\alpha
_{1}{S}_{x}-\alpha _{2}{S}_{y},\alpha _{2}{S}_{x}+\alpha
_{1}S_{y})$ is the SO vector potential. For $S=0$, the eigenstates
are Landau levels for a charge 2$e$ particle in the magnetic field
$B$, analogous to the spinless problem \cite{Kawabata}. For $S=1$,
eigenstates of $\mathcal{H}$ and $P_{1}$ in general require a
numerical solution, although analytic solutions exist when either
$\alpha_1$ or $\alpha_2$ equals zero \cite{Lyanda1}. An analytic
solution is also found for $\alpha_1$, $\alpha_2 \neq 0$, when
$\ell_{B} < \lambda_{so}$, where
$\lambda_{so}=(2\alpha_{1(2)}m^*/\hbar^{2})^{-1}$ is the distance
over which spin rotates appreciably (if $\ell > \lambda_{so}$) or
dephases (if $\ell < \lambda_{so}$) due to spin AB-like effects.
Performing a unitary transformation $\mathcal{H}\rightarrow
{\tilde{\mathcal{H}}}=U^{\dagger }\mathcal{H}U$, with $U=\exp
{(-ie\mathbf{A}_{S}\cdot \mathbf{r})}$, and expanding in
coordinates, we find ${\tilde{\mathcal{H}}}=\frac{\hbar
^{2}}{2m^*}(\mathbf{k}-2e\mathbf{A}_{em}+S_{z}\mathbf{a})^{2}$,
where $\mathbf{a}=H_{ \mathrm{eff}}\,\mathbf{r}\times
{\mathbf{\hat{z}}/(2\hbar^2),}$ and $H_{\mathrm{eff} }=2(\alpha
_{2}^{2}-\alpha _{1}^{2}){m^*}^{2}/e\hbar ^{3}$ is the effective
SO field. $P_{1}$ can then be block-diagonalized for each $m$
($m=0,\pm1$) using the Landau basis for particles with charge 2$e$
in the magnetic field $B-mH_{\mathrm{eff}}$. Thus, the effect of
$\mathbf{\Omega}_{D1}$ and $\mathbf{\Omega}_{R}$ is to produce
AB-like spin phases
\cite{Aronov,Stone,Mirlin,Iordanskii94,Lyanda98,Aleiner}. Higher
expansion terms to $\tilde{\mathcal{H}}$ describe spin flip
processes and can be taken into account by introducing a spin
relaxation time $\tau _{so}$ and its corresponding field scale
$H_{so} = \hbar \tau/(2e \ell^2 \tau_{so})$. The resulting quantum
interference contribution takes the form \cite{Lyanda1}


\begin{equation}\label{Cond}
\Delta \sigma (B)=-\frac{e^{2}}{4\pi ^{2}\hbar }\left[
\sum_{m=-1,0,1}C(x_{1m},f_{1m})-C(x_{00},f_{00})\right]
\end{equation}
\noindent where $x_{Sm}=(B-mH_{\mathrm{eff}}^{{}})/H_{tr}$
describes the AB dephasing in $H_{\mathrm{eff}}^{{}}$,
$C(x,f_{Sm})=x\sum_{N=0}^{\infty}\frac{P_{N}^{3}(f_{Sm})}{1-P_{N}(f_{Sm})}$,
$P_{N}(f_{Sm})=y\int_{0}^{\infty}\exp(-yf_{Sm}t-t^{2}/2)L_{N}(t^{2})dt$,
$L_{N}(z)$ are Laguerre polynomials, $y=(2/|x|)^{1/2}$, and
$H_{tr} = \hbar/(2e \ell^2)$. The dephasing factors $f_{Sm}$ are
given by $f_{1\pm 1}=(1+(H_{\varphi
}+H_{so})/H_{tr});\,f_{00}=(1+H_{\varphi
}/H_{tr});\,f_{10}=(1+(H_{\varphi }+2H_{so})/H_{tr})$, where
$H_{\varphi} = \hbar/(4e L_{\varphi}^2)$ and $L_{\varphi}$ is the
phase breaking length.
Equation (\ref{Cond}) does not include all $B$-dependent
interference terms, notably excluding Cooper-channel contributions
due to electron-electron interactions \cite{Altshuler1} and a
reduction of WL due to electron diffraction effects
\cite{Gasparyan85}. Also, in an attempt to capture the effects of
cubic terms on $H_{\mathrm{eff}}$ and $H_{so}$, we introduce an
effective vector potential $\mathbf{ A}_{S}^{\ast
}=\mathbf{A}_{S}+\gamma \frac{m^*}{e\hbar
^{2}}({k_y}^{2},-{k_x}^{2})\sim \mathbf{A}_{S}+\gamma
\frac{m^*}{2e\hbar ^{2}}(k^2,-k^2)$ which leads to an effective SO
field,
\begin{equation}\label{Heff} H_{\mathrm{eff}}^{\ast }=2(\alpha
_{2}^{2}-\alpha _{1}^{2}+2\pi n\alpha
_{1}\gamma - \pi ^{2}\gamma ^{2}n^{2}){m^*}^{2}/e\hbar ^{3}.
\end{equation}
\noindent Equation~(\ref{Cond}) is applicable when $B
>  H^*_{\mathrm{eff}}$
(see Fig~2). We have confirmed that fitting only to data where $B
>  H^*_{\mathrm{eff}}$ gives, within error bars, the same results
as fitting over the entire measured range of $B$.
Modification of the commutator $[\mathbf{k}+2e\mathbf{A}_{S}^{\ast
},{\mathbf{r]}}$ by $\mathbf{A}_{S}^{\ast }$ induces spin flipping
terms $\sim \gamma k^3/4$ in the transformed Hamiltonian
${\tilde{\mathcal{H}}}^{\ast }$.  The corresponding $H_{so}^{\ast
}=\frac{1}{36}\pi ^{2}{m^*}^{2}\gamma ^{2}n^{2}/e\hbar$, using its
expression in the diffusive regime.

\begin{figure}[t]
\center \label{fig2} \includegraphics[width=2.4in]{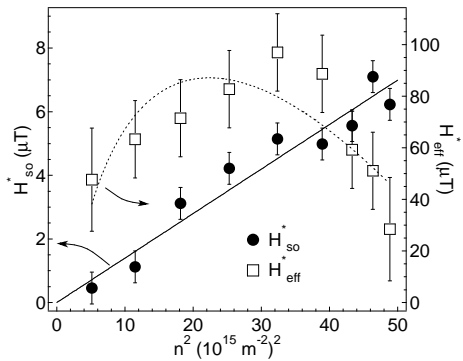}
\caption{\footnotesize {Spin-orbit effective fields, $H^*_{so}$
(filled circles) and $H^*_{\mathrm{eff}}$ (open squares), as
extracted using Eq.~(\ref{Cond}), plotted as a function of sheet
density squared. The best fit of Eq. (\ref{Heff}) to
$H^*_{\mathrm{eff}}$ (dotted curve) is used to extract $\gamma$,
$\alpha_1$ and $\alpha_2$.  Alternatively, the best linear fit to
$H^*_{so}$ (solid line) is used to extract $\gamma$.}}
\end{figure}

We now turn to a discussion of the experiment. Samples on three
separate heterostructure materials all showed qualitatively
similar behavior.  The sample for which data is presented consists
of a GaAs/AlGaAs heterostructure grown in the [001] direction with
double $\delta$-doping layers set back 143~\AA\ and 161~\AA\ from
the 2DEG and a total distance of 349~\AA\ from the surface to the
2DEG. A 200~$\mu$m wide Hall bar with 700~$\mu$m between voltage
probes was patterned by wet etching. A lithographically defined
Cr/Au top gate was used to control density and mobility in the
Hall bar over the range $n=$~1.4-7.0~$\times 10^{15}$~m$^{-2}$ and
$\mu=$~3.6-31~m$^2$/Vs. Measurements were made in a $^3$He
cryostat at temperature $T = 300$~mK using ac lock-in techniques
with bias currents ranging from 50 to 500~nA.
\begin{figure}[t]
\center \label{fig3} \includegraphics[width=2.8in]{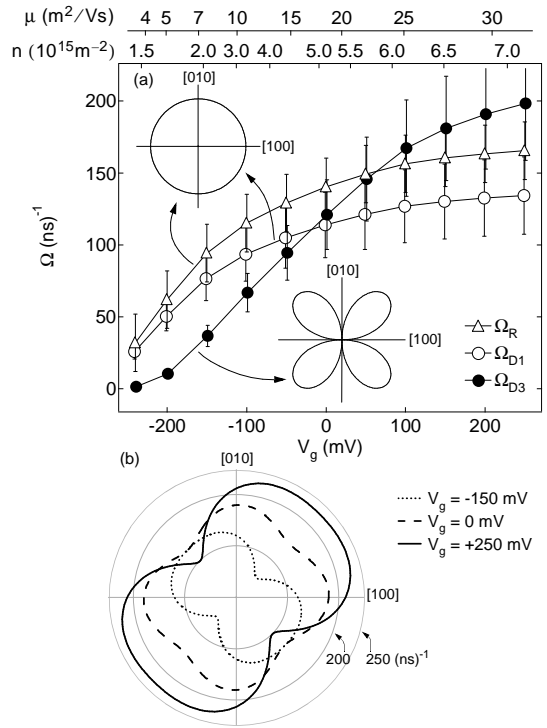}
\caption{\footnotesize {(a) Magnitudes of isotropic linear
Dresselhaus ($\Omega_{D1}$) and Rashba ($\Omega_R$) terms, and
nonisotropic cubic Dresselhaus  ($\Omega_{D3}$) term as functions
of gate voltage, $V_g$, density, $n$, and mobility, $\mu$. Insets
show theoretical dependence on momentum direction for the three
terms, indicating that the linear terms are isotropic, while the
cubic term has a four-fold symmetry and is highly anisotropic.
Maximum magnitude (when $\phi = (j+\frac{1}{4})\pi$) is shown for
the anisotropic ($\Omega_{D3}$) term. (b) Angular variation of
$\Omega$, the magnitude of the total SO precession vector at $V_g
= -150$ mV (dotted), 0 mV (dashed), and 250 mV (solid),
corresponding to densities of 2.3, 5.0, and $7.0 \times 10^{15}$
m$^{-2}$ respectively. }}
\end{figure}
Figure~1(a) shows the longitudinal magnetoconductance as a
function of $V_{g}$. A crossover from pure WL (Fig.~1(a), inset)
at $V_{g}=-240$~mV to essentially pure AL at $V_{g}=+250$~mV is
observed. This crossover demonstrates that a gate can be used to
control SO over a wide range, as pure WL corresponds to negligible
SO rotations within the phase coherence length $L_{\varphi }$,
while AL corresponds to spin rotations $\gtrsim 2\pi$. The solid
curves in Fig.~1(a) are fits of Eq.~(\ref{Cond}) with three free
parameters, $H_{\phi }$, $H^*_{so}$, and $H^*_{\mathrm{eff}}$.
$H_{tr}$ is fixed at each gate voltage by measured values of
density and mobility.
Figure~2 shows extracted parameters $H^*_{so}$ and
$H^*_{\mathrm{eff}}$ as a function of $n^2$.  $H^*_{so}$ is well
described by the predicted linear dependence on $n^2$, with a best
fit (Fig.~2, solid line) giving $\gamma =31\pm 3$~eV\AA$^{3}$ with
zero y-intercept (see Eq.~(\ref{Omega3Eqn})). The density
dependence of $H_{\mathrm{eff}}^{*}$ is well described by
Eq.~(\ref{Heff}), (Fig.~2, dotted curve), giving fit parameters
$\gamma =28\pm 4$~eV\AA$^{3}$, $\alpha _{1}=4\pm 1$~meV\AA\ and
$\alpha_{2}=5\pm 1$~meV\AA. In this way, the three SO parameters
$\alpha_1$, $\alpha_2$, and $\gamma$ are separately obtained from
transport measurements by explicitly making use of the density
dependence of $H^*_{\mathrm{eff}}$ and $H^*_{so}$. Extracted
values of $H_{\varphi}$ correspond to dephasing times in the range
$\tau_{\varphi} \sim 0.1$-$1.0$~ns at 300~mK, which decrease by
more than an order of magnitude as temperature is increased to
2.5~K. Within the error bars, $H_{so}^*$ and $H_{\mathrm{eff}}^*$
do not depend on temperature over this temperature range.
Figure~3(a) displays the magnitudes of the three spin-orbit terms
as functions of $V_g$, $n$, and $\mu$, determined using
Eq.~(\ref{Omegas}) and the extracted values of $\alpha_1$,
$\alpha_2$, and $\gamma$.  Plotted are values along the [110]
direction, $\phi\equiv \tan ^{-1}(k_y/k_x)=\frac{\pi}{4}$, where
$\Omega_{D3}$ is maximum.
The total spin precession rate, $\Omega$, is plotted as a function
of the direction, $\phi$, of the electron momentum in Fig.~3(b).
While for most directions $\Omega$ is an increasing function of
density, it is seen to decrease with increasing density near
$\phi= \frac{3\pi}{4}$ and $\frac{7\pi}{4}$. The linear
Dresselhaus and Rashba terms ($\Omega_{D1}$ and $\Omega_{R}$) are
of comparable magnitude to each other for all densities and in all
directions. Near $\phi= \frac{j\pi}{2}$ ($j$ an integer),
$\Omega_{D3} \ll \Omega_{D1}, \Omega_{R}$ and the SO is controlled
by the linear terms. For $\phi$ near $\frac{(2j+1)\pi}{4}$, the
cubic term becomes comparable to or even exceeds (at high
densities) the linear terms. Depending on $\phi$, the linear and
cubic terms either add ($\phi\sim \frac{\pi}{4},\frac{5\pi}{4}$)
or subtract ($\phi\sim \frac{3\pi}{4}, \frac{7\pi}{4}$).
The extracted values for $\gamma$ ($31\pm 3$~eV\AA$^{3}$ using
$H_{so}^*$, $28\pm 4$~eV\AA$^{3}$ using $H_{\mathrm{eff}}^*$) are
in good agreement with the value 27.5 eV\AA $^{3}$ from band
structure calculations \cite{Knap96,Pfeffer99}. Values of
$\Omega$ are $\sim3-8$ times smaller than previously
measured using Shubnikov-deHaas oscillations
\cite{Ramvall97}, with corresponding theory \cite{Ramvall97,
Pfeffer99} lying roughly between the experimental ranges. We
note, however, that the values are sample dependent.   Estimates for
$\alpha_1$ give values for $\langle k_{z}^{2}\rangle$ that
correspond to a wave function width of $\sim 10$~nm in the
$\hat{z}$ direction, which is also reasonable. The extracted
$\alpha _{2}$ corresponds to a uniform
electric field $E\sim10$ MV/m, using $\alpha _{2}=\alpha _{0}eE$
and a value of
$\alpha_{0}=5.33$~\AA$^2$ from a
$\mathbf{k\cdot p}$ model \cite{Knap96,Pfeffer99}.
Previously existing models for WL/AL \cite{AALKh81,Iordanskii94,
Knap96} provide fits to the data that appear qualitatively
reasonable, giving values for $H_{so}$ that are $\sim 5$ times
higher than those found using Eq.~(\ref{Cond}). However, these
fits also lead to the unphysical result that $\tau_{so}<\tau$.
Such unphysical results are not surprising given that, for $V_g >
-50$ mV, the SO length, $v_{F}/\langle\Omega\rangle$, is less than
$\ell$, while theory \cite{AALKh81, Iordanskii94, Knap96} assumes
diffusive spin evolution $\ell \ll \lambda_{so},L_{\varphi}$.  We
note that a theory for arbitrarily strong SO coupling
\cite{Lyanda98} may also be used to fit this data by including $B$
via $L_{\varphi}$. This approach yields values for $\Omega _{D3}$
and $\Omega _{D1}$ consistent with Eq.~(\ref{Cond}) to within a
factor of $\sim 3$, but does not separate $\Omega _{D1}$ and
$\Omega _{R}$ terms.
We thank I.~Aleiner, H.~Bruus and S. Studenikin for illuminating
discussions and F.~Mancoff for device fabrication.  This work was
supported in part by DARPA-QuIST, DARPA-SpinS, ARO-MURI, and
NSF-NSEC. We also acknowledge support from ONR and NSA (Y.~L.-G.),
NDSEG (J.~B.~M.) and the Harvard Society of Fellows (D.G.-G). Work
at UCSB was supported by QUEST, an NSF Science and Technology
Center.


\small

\begin{thebibliography}{39}
\expandafter\ifx\csname
natexlab\endcsname\relax\def\natexlab#1{#1}\fi
\expandafter\ifx\csname bibnamefont\endcsname\relax
   \def\bibnamefont#1{#1}\fi
\expandafter\ifx\csname bibfnamefont\endcsname\relax
   \def\bibfnamefont#1{#1}\fi
\expandafter\ifx\csname citenamefont\endcsname\relax
   \def\citenamefont#1{#1}\fi
\expandafter\ifx\csname url\endcsname\relax
   \def\url#1{\texttt{#1}}\fi
\expandafter\ifx\csname
urlprefix\endcsname\relax\def\urlprefix{URL }\fi
\providecommand{\bibinfo}[2]{#2}
\providecommand{\eprint}[2][]{\url{#2}}
\bibitem[{\citenamefont{Wolf et~al.}(2001)}]{Wolf}
\bibinfo{author}{\bibfnamefont{S.~A.} \bibnamefont{Wolf}} \bibnamefont{et~al.},
   \bibinfo{journal}{Science} \textbf{\bibinfo{volume}{294}},
   \bibinfo{pages}{1488} (\bibinfo{year}{2001}).
\bibitem[{\citenamefont{Awschalom et~al.}(2002)}]{AwschalomBook}
\bibinfo{author}{\bibfnamefont{D.~D.} \bibnamefont{Awschalom}}
   \bibnamefont{et~al.}, \emph{\bibinfo{title}{Semiconductor Spintronics and
   Quantum Computation}} (\bibinfo{publisher}{Springer-Verlag},
   \bibinfo{year}{2002}).
\bibitem[{\citenamefont{Datta and Das}(1990)}]{Datta90}
\bibinfo{author}{\bibfnamefont{S.}~\bibnamefont{Datta}} \bibnamefont{and}
   \bibinfo{author}{\bibfnamefont{B.}~\bibnamefont{Das}},
   \bibinfo{journal}{Appl.\ Phys.\ Lett.} \textbf{\bibinfo{volume}{56}},
   \bibinfo{pages}{665} (\bibinfo{year}{1990}).

\bibitem[{\citenamefont{Aronov and Lyanda-Geller}(1993)}]{Aronov}
\bibinfo{author}{\bibfnamefont{A.~G.} \bibnamefont{Aronov}} \bibnamefont{and}
   \bibinfo{author}{\bibfnamefont{Y.~B.} \bibnamefont{Lyanda-Geller}},
   \bibinfo{journal}{Phys. Rev. Lett.} \textbf{\bibinfo{volume}{70}},
   \bibinfo{pages}{343} (\bibinfo{year}{1993}).

\bibitem[{\citenamefont{Koga et~al.}(2002)}]{Koga}
\bibinfo{author}{\bibfnamefont{T.}~\bibnamefont{Koga}} \bibnamefont{et~al.},
   \bibinfo{journal}{Phys. Rev. Lett.} \textbf{\bibinfo{volume}{88}},
   \bibinfo{pages}{126601} (\bibinfo{year}{2002}).

\bibitem[{\citenamefont{Kiselev and Kim}(2001)}]{Kiselev}
\bibinfo{author}{\bibfnamefont{A.}~\bibnamefont{Kiselev}} \bibnamefont{and}
   \bibinfo{author}{\bibfnamefont{K.}~\bibnamefont{Kim}},
   \bibinfo{journal}{Appl. Phys. Lett.} \textbf{\bibinfo{volume}{78}},
   \bibinfo{pages}{775} (\bibinfo{year}{2001}).

~~~~~\bibitem[{\citenamefont{Hikami et~al.}(1980)}]{Hikami80}
\bibinfo{author}{\bibfnamefont{S.}~\bibnamefont{Hikami}} \bibnamefont{et~al.},
   \bibinfo{journal}{Prog. Theor. Phys.} \textbf{\bibinfo{volume}{63}},
   \bibinfo{pages}{707} (\bibinfo{year}{1980}).


\bibitem[{\citenamefont{Altshuler et~al.}(1981)}]{AALKh81}
\bibinfo{author}{\bibfnamefont{B.}~\bibnamefont{Altshuler}}
   \bibnamefont{et~al.}, \bibinfo{journal}{JETP} \textbf{\bibinfo{volume}{54}},
   \bibinfo{pages}{411} (\bibinfo{year}{1981}).
\bibitem[{\citenamefont{Bergmann}(1984)}]{Bergmann}
\bibinfo{author}{\bibfnamefont{G.}~\bibnamefont{Bergmann}},
   \bibinfo{journal}{Phys. Rep.} \textbf{\bibinfo{volume}{107}},
   \bibinfo{pages}{1} (\bibinfo{year}{1984}).
\bibitem[{\citenamefont{Dresselhaus et~al.}(1992)}]{Dresselhaus92}
\bibinfo{author}{\bibfnamefont{P.~D.} \bibnamefont{Dresselhaus}}
   \bibnamefont{et~al.}, \bibinfo{journal}{Phys.\ Rev.\ Lett.}
   \textbf{\bibinfo{volume}{68}}, \bibinfo{pages}{106} (\bibinfo{year}{1992}).
\bibitem[{\citenamefont{Mathur and Stone}(1992)}]{Stone}
\bibinfo{author}{\bibfnamefont{H.}~\bibnamefont{Mathur}} \bibnamefont{and}
   \bibinfo{author}{\bibfnamefont{A.~D.} \bibnamefont{Stone}},
   \bibinfo{journal}{Phys. Rev. Lett.} \textbf{\bibinfo{volume}{68}},
   \bibinfo{pages}{2964} (\bibinfo{year}{1992}).
\bibitem[{\citenamefont{Lyanda-Geller and Mirlin}(1994)}]{Mirlin}
\bibinfo{author}{\bibfnamefont{Y.~B.} \bibnamefont{Lyanda-Geller}}
   \bibnamefont{and} \bibinfo{author}{\bibfnamefont{A.~D.}
   \bibnamefont{Mirlin}}, \bibinfo{journal}{Phys. Rev. Lett.}
   \textbf{\bibinfo{volume}{72}}, \bibinfo{pages}{1894} (\bibinfo{year}{1994}).
\bibitem[{\citenamefont{Iordanskii et~al.}(1994)}]{Iordanskii94}
\bibinfo{author}{\bibfnamefont{S.~V.} \bibnamefont{Iordanskii}}
   \bibnamefont{et~al.}, \bibinfo{journal}{JETP\ Lett.}
   \textbf{\bibinfo{volume}{60}}, \bibinfo{pages}{206} (\bibinfo{year}{1994}).
\bibitem[{\citenamefont{Lyanda-Geller}(1998)}]{Lyanda98}
\bibinfo{author}{\bibfnamefont{Y.}~\bibnamefont{Lyanda-Geller}},
   \bibinfo{journal}{Phys.\ Rev.\ Lett.} \textbf{\bibinfo{volume}{80}},
   \bibinfo{pages}{4273} (\bibinfo{year}{1998}).
\bibitem[{\citenamefont{Aleiner and Fal'ko}(2001)}]{Aleiner}
\bibinfo{author}{\bibfnamefont{I.~L.} \bibnamefont{Aleiner}} \bibnamefont{and}
   \bibinfo{author}{\bibfnamefont{V.~I.} \bibnamefont{Fal'ko}},
   \bibinfo{journal}{Phys. Rev. Lett.} \textbf{\bibinfo{volume}{87}},
   \bibinfo{pages}{256801} (\bibinfo{year}{2001}).
\bibitem[{\citenamefont{Zumb\"{u}hl et~al.}(2002)}]{Zumbuhl}
\bibinfo{author}{\bibfnamefont{D.~M.}~\bibnamefont{Zumb\"{u}hl}}
   \bibnamefont{et~al.}, \bibinfo{journal}{Phys.\ Rev.\ Lett.} \textbf{\bibinfo{volume}{89}},
   \bibinfo{pages}{276803} (\bibinfo{year}{2002}).

\bibitem[{\citenamefont{Abrahams et~al.}(1979)}]{Anderson}
\bibinfo{author}{\bibfnamefont{E.}~\bibnamefont{Abrahams}}
   \bibnamefont{et~al.}, \bibinfo{journal}{Phys. Rev. Lett.}
   \textbf{\bibinfo{volume}{42}}, \bibinfo{pages}{673} (\bibinfo{year}{1979}).
\bibitem[{\citenamefont{Altshuler and Aronov}(1985)}]{Altshuler1}
\bibinfo{author}{\bibfnamefont{B.~L.} \bibnamefont{Altshuler}}
   \bibnamefont{and} \bibinfo{author}{\bibfnamefont{A.~G.}
   \bibnamefont{Aronov}}, \emph{\bibinfo{title}{Electron-Electron Interactions
   in Disordered Systems, ed by A. L. Efros and M. Pollak}}
   (\bibinfo{publisher}{North Holland, Amsterdam}, \bibinfo{year}{1985}),
   p.~\bibinfo{pages}{11}.
\bibitem[{\citenamefont{Gorkov et~al.}(1980)}]{Gorkov}
\bibinfo{author}{\bibfnamefont{L.}~\bibnamefont{Gorkov}} \bibnamefont{et~al.},
   \bibinfo{journal}{JETP Lett.} \textbf{\bibinfo{volume}{30}},
   \bibinfo{pages}{228} (\bibinfo{year}{1980}).
\bibitem[{\citenamefont{Lyanda-Geller}(2002)}]{Lyanda1}
\bibinfo{author}{\bibfnamefont{Y.}~\bibnamefont{Lyanda-Geller}}
   (\bibinfo{year}{2002}), \bibinfo{note}{unpublished}.
\bibitem[{\citenamefont{Gasparyan and Zyuzin}(1985)}]{Gasparyan85}
\bibinfo{author}{\bibfnamefont{V.~M.} \bibnamefont{Gasparyan}}
   \bibnamefont{and} \bibinfo{author}{\bibfnamefont{A.~Y.}
   \bibnamefont{Zyuzin}}, \bibinfo{journal}{Sov.\ Phys.\ Solid State}
   \textbf{\bibinfo{volume}{27}}, \bibinfo{pages}{999} (\bibinfo{year}{1985}).
\bibitem[{\citenamefont{Cassam-Chenai and Shapiro}(1994)}]{REF:Shapiro}
\bibinfo{author}{\bibfnamefont{A.}~\bibnamefont{Cassam-Chenai}}
   \bibnamefont{and} \bibinfo{author}{\bibfnamefont{B.}~\bibnamefont{Shapiro}},
   \bibinfo{journal}{J. Phys. I} \textbf{\bibinfo{volume}{4}},
   \bibinfo{pages}{1527} (\bibinfo{year}{1994}).
\bibitem[{\citenamefont{Kawabata}(1984)}]{Kawabata}
\bibinfo{author}{\bibfnamefont{A.}~\bibnamefont{Kawabata}},
   \bibinfo{journal}{J. Phys. Soc. Jpn.} \textbf{\bibinfo{volume}{53}},
   \bibinfo{pages}{3540} (\bibinfo{year}{1984}).
\bibitem[{\citenamefont{Zduniak et~al.}(1997)}]{Zduniak97}
\bibinfo{author}{\bibfnamefont{A.}~\bibnamefont{Zduniak}} \bibnamefont{et~al.},
   \bibinfo{journal}{Phys. Rev. B} \textbf{\bibinfo{volume}{56}},
   \bibinfo{pages}{1996} (\bibinfo{year}{1997}).
\bibitem[{\citenamefont{Hansen et~al.}(1993)}]{Taborski}
\bibinfo{author}{\bibfnamefont{J.~E.} \bibnamefont{Hansen}}
   \bibnamefont{et~al.}, \bibinfo{journal}{Phys. Rev. B}
   \textbf{\bibinfo{volume}{47}}, \bibinfo{pages}{16040} (\bibinfo{year}{1993}).
\bibitem[{\citenamefont{Ramvall et~al.}(1997)}]{Ramvall97}
\bibinfo{author}{\bibfnamefont{P.}~\bibnamefont{Ramvall}} \bibnamefont{et~al.},
   \bibinfo{journal}{Phys. Rev. B} \textbf{\bibinfo{volume}{55}},
   \bibinfo{pages}{7160} (\bibinfo{year}{1997}).
\bibitem[{\citenamefont{Koga et~al.}(2002{\natexlab{b}})}]{Koga2}
\bibinfo{author}{\bibfnamefont{T.}~\bibnamefont{Koga}} \bibnamefont{et~al.},
   \bibinfo{journal}{Phys. Rev. Lett.} \textbf{\bibinfo{volume}{89}},
   \bibinfo{pages}{046801} (\bibinfo{year}{2002}{\natexlab{b}}).
\bibitem[{\citenamefont{Nitta et~al.}(1997)}]{Nitta97}
\bibinfo{author}{\bibfnamefont{J.}~\bibnamefont{Nitta}} \bibnamefont{et~al.},
   \bibinfo{journal}{Phys. Rev. Lett.} \textbf{\bibinfo{volume}{78}},
   \bibinfo{pages}{1335} (\bibinfo{year}{1997}).
\bibitem[{\citenamefont{Sch\"apers et~al.}(1998)}]{Schapers98}
\bibinfo{author}{\bibfnamefont{T.}~\bibnamefont{Sch\"apers}}
   \bibnamefont{et~al.}, \bibinfo{journal}{J.\ Appl.\ Phys.}
   \textbf{\bibinfo{volume}{83}}, \bibinfo{pages}{4324} (\bibinfo{year}{1998}).
\bibitem[{\citenamefont{Heida et~al.}(1998)}]{Heida98}
\bibinfo{author}{\bibfnamefont{J.~P.} \bibnamefont{Heida}}
   \bibnamefont{et~al.}, \bibinfo{journal}{Phys. Rev. B}
   \textbf{\bibinfo{volume}{57}}, \bibinfo{pages}{11911} (\bibinfo{year}{1998}).
\bibitem[{\citenamefont{Schultz et~al.}(1996)}]{Schultz96}
\bibinfo{author}{\bibfnamefont{M.}~\bibnamefont{Schultz}} \bibnamefont{et~al.},
   \bibinfo{journal}{Semicond. Sci. Technol.} \textbf{\bibinfo{volume}{11}},
   \bibinfo{pages}{1168} (\bibinfo{year}{1996}).
\bibitem[{\citenamefont{Papadakis et~al.}(2001)}]{Papadakis01}
\bibinfo{author}{\bibfnamefont{S.~J.} \bibnamefont{Papadakis}}
   \bibnamefont{et~al.}, \bibinfo{journal}{Physica E}
   \textbf{\bibinfo{volume}{9}}, \bibinfo{pages}{31} (\bibinfo{year}{2001}).
\bibitem[{\citenamefont{Winkler et~al.}(2002)}]{Winkler02}
\bibinfo{author}{\bibfnamefont{R.}~\bibnamefont{Winkler}} \bibnamefont{et~al.},
   \bibinfo{journal}{Phys. Rev. B} \textbf{\bibinfo{volume}{65}},
   \bibinfo{pages}{155303} (\bibinfo{year}{2002}).
\bibitem[{\citenamefont{Lu et~al.}(1998)}]{Lu98}
\bibinfo{author}{\bibfnamefont{J.~P.} \bibnamefont{Lu}} \bibnamefont{et~al.},
   \bibinfo{journal}{Phys. Rev. Lett.} \textbf{\bibinfo{volume}{81}},
   \bibinfo{pages}{1282} (\bibinfo{year}{1998}).
\bibitem[{\citenamefont{Jusserand et~al.}(1995)}]{Jusserand95}
\bibinfo{author}{\bibfnamefont{B.}~\bibnamefont{Jusserand}}
   \bibnamefont{et~al.}, \bibinfo{journal}{Phys. Rev. B}
   \textbf{\bibinfo{volume}{51}}, \bibinfo{pages}{4707} (\bibinfo{year}{1995}).
\bibitem[{\citenamefont{Knap et~al.}(1996)}]{Knap96}
\bibinfo{author}{\bibfnamefont{W.}~\bibnamefont{Knap}} \bibnamefont{et~al.},
   \bibinfo{journal}{Phys.\ Rev.\ B} \textbf{\bibinfo{volume}{53}},
   \bibinfo{pages}{628} (\bibinfo{year}{1996}).
\bibitem[{\citenamefont{Pfeffer}(1999)}]{Pfeffer99}
\bibinfo{author}{\bibfnamefont{P.}~\bibnamefont{Pfeffer}},
   \bibinfo{journal}{Phys. Rev. B} \textbf{\bibinfo{volume}{59}},
   \bibinfo{pages}{15902} (\bibinfo{year}{1999}).
\end{thebibliography}
\end{document}